\documentclass[aps, reprint]{revtex4-2}

\usepackage{graphicx}% Include figure files
\usepackage{amsmath,amssymb}

\usepackage{dcolumn}

\usepackage[dvipsnames]{xcolor}
\usepackage[%  
    colorlinks=true,
    citecolor = blue,
    linkcolor= blue,
    urlcolor=blue
]{hyperref}
\usepackage{hyperref} %for linking github repos
 
\newcommand{\fig}{Fig.~}
\newcommand{\suppfig}{Supplementary Fig.~}

\newcommand{\pself}{p_\textrm{s}}
\newcommand{\pnei}{p_\textrm{r}}
\newcommand{\pext}{p_\textrm{ext}}
\newcommand{\rad}{k}
\newcommand{\bp}{m}
\newcommand{\expo}{\tau}
\newcommand{\prw}{p_\textrm{rw}} 
\newcommand{\coal}{C}
\newcommand{\av}{S}
\newcommand{\suppsec}{Appendix~}

\DeclareMathOperator{\expectedvalue}{\mathbb{E}} 
\newcommand{\EX}[1]{\expectedvalue \left[ #1 \right]}

\makeatletter
\def\maketitle{
\@author@finish
\title@column\titleblock@produce
\suppressfloats[t]}
\makeatother

\begin{document}

\title{Topology-dependent coalescence controls scaling exponents in finite networks}

\author{Roxana Zeraati $^{1,2,*}$, Victor Buend\'{\i}a$^{3,2,*}$, Tatiana A. Engel $^{4}$, Anna Levina $^{3,2,5}$\\}

\affiliation{
$^1$ International Max Planck Research School for the Mechanisms of Mental Function and Dysfunction, University of T\"ubingen, T\"ubingen, Germany\\
$^2$ Max Planck Institute for Biological Cybernetics, T\"ubingen, Germany\\
$^3$ Department of Computer Science, University of T\"ubingen, T\"ubingen, Germany\\
$^4$ Cold Spring Harbor Laboratory, Cold Spring Harbor, NY, USA\\
$^5$ Bernstein Center for Computational Neuroscience T\"ubingen, T\"ubingen, Germany\\
{\small $^{*}$ These authors contributed equally to this work.}\\
{\small $^\dagger$Corresponding author e-mail: anna.levina@uni-tuebingen.de}}

\begin{abstract}

Multiple studies of neural avalanches across different data modalities led to the prominent hypothesis that the brain operates near a critical point.
The observed exponents often indicate the mean-field directed-percolation universality class, leading to the fully-connected or random network models to study the avalanche dynamics.
However, the cortical networks have distinct non-random features and spatial organization that is known to affect the critical exponents.
Here we show that distinct empirical exponents arise in networks with different topology and depend on the network size.
In particular, we find apparent scale-free behavior with mean-field exponents appearing as quasi-critical dynamics in structured networks.
This quasi-critical dynamics cannot be easily discriminated from an actual critical point in small networks.
We find that the local coalescence in activity dynamics can explain the distinct exponents.
Therefore, both topology and system size should be considered when assessing criticality from empirical observables.

\end{abstract}

\maketitle

\begin{figure*}[t]
    \centering
     \includegraphics[trim=0 0 0 0, clip, width = 1\linewidth]{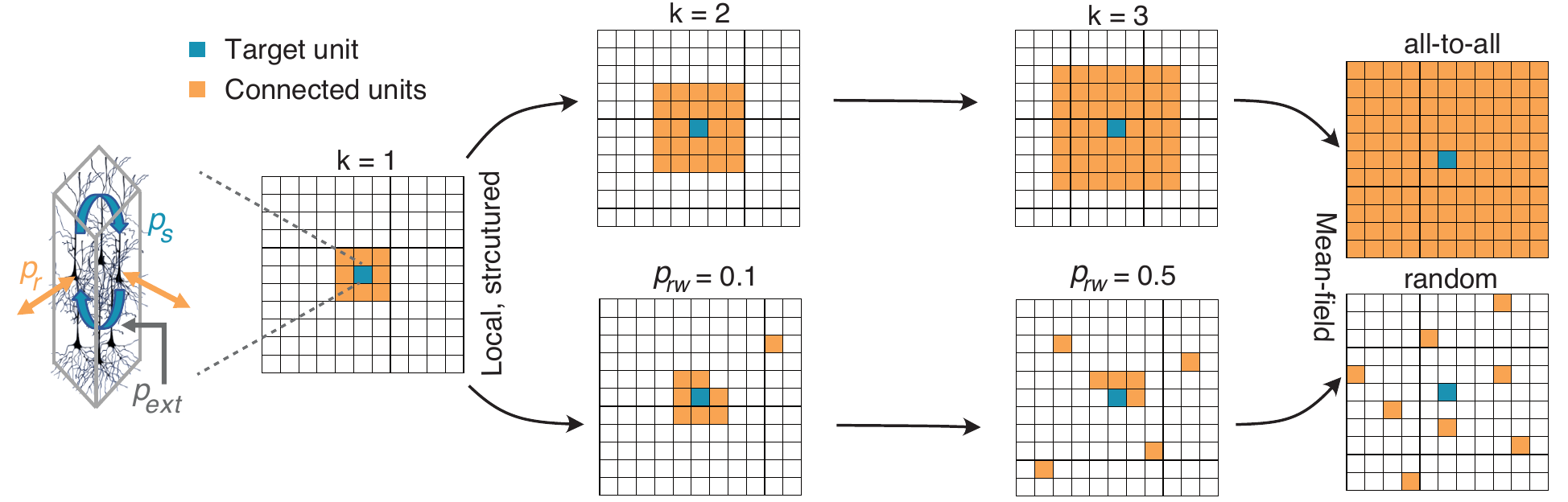}
    \caption{{Branching network model with different types of spatial connectivity.} 
    Each unit in the network represents a cortical column (left) that can be excited by the self-excitation (probability $\pself$, blue arrows), its neighboring units (probability $\pnei$, orange arrows), and the external input (probability $\pext$, gray arrow).
    We consider networks with connectivity structures ranging from local, structured (left) to mean-field (random or all-to-all, right) generated via two pathways: by increasing the connectivity radius $\rad$ (top) or rewiring local connections to random with the rewiring probability $\prw$ (bottom).
    }
    \label{fig:model_conn}
\end{figure*}

Brain activity displays a plethora of different dynamical states, including bursts, oscillations, and irregular activity. 
In particular, neural activity exhibits spatiotemporal patterns compatible with the dynamics of a system close to a second-order phase transition. Operating at this regime has been linked to optimal information processing~\cite{beggs_the_2008,munoz_colloquium_2018}, maximal dynamic range and sensitivity to stimulus~\cite{kinouchi_optimal_2006,zierenberg_tailored_2020}, longer timescales during selective attention~\cite{zeraati_attentional_2021}, and better stimuli discrimination~\cite{bertschinger2004, tomen_marginally_2014}. 
Neural network models demonstrated that short- and long-term synaptic plasticity can self-organize brain dynamics towards a critical point~\cite{zeraati_self-organization_2021,buendia_feedback_2020}.

To assess whether the brain dynamics is critical, the activity propagation between the neurons is often mapped to the branching process~\cite{beggs_neuronal_2003,kinouchi_optimal_2006,beggs_the_2008,petermann_spontaneous_2009,pasquale_self-organization_2008,poil_avalanche_2008,bellay_irregular_2015,palva_neuronal_2013,zierenberg_description_2020}.
This mapping was motivated by in-vitro observation of outbursts of neural activity known as \emph{neuronal avalanches} with sizes and durations following power-law distributions with exponents $\tau=1.5$ and $\alpha=2$ correspondingly~\cite{beggs_neuronal_2003}, as expected from the branching process at the critical point.
The branching process dynamics is fully characterized by a branching parameter $\bp$~\cite{harris_theory_2002}, with $\bp = 1$ being the critical value, where each neuron on average activates one other neuron creating a fluctuation-driven regime.
From the statistical mechanics point of view, the critical transition at $\bp=1$ belongs to the \emph{mean-field directed percolation} (MF-DP) universality class, a non-equilibrium phase transition separating absorbing and active phases \cite{hinrichsen_non-equilibrium_2000,henkel_non-equilibrium_2008a}. 
In this universality class, the avalanche-size distribution has a power-law exponent of $1.5$.

However, mapping the neural activity propagation to the branching process neglects the role of underlying network topology in shaping the dynamics.
The branching process assumes non-overlapping spreading of the activity.
In biological neural networks, in contrast, each neuron can be simultaneously excited by multiple sources. 
This phenomenon, known as \textit{coalescence} renders the independence assumption invalid and reduces the effective branching parameter of the system since some active neurons cannot trigger spikes in already excited neighbors~\cite{zierenberg_description_2020}.
These effects are particularly severe in structured networks, for example as observed in primate cortex~\cite{mountcastle_columnar_1997,casanova_modular_2019, smith_spatial_2008,safavi_nonmonotonic_2018,rosenbaum_spatial_2017}. 
Additionally, the theory of critical phenomena predicts that structured, finite-dimensional network topology affects variables such as critical exponents~\cite{munoz_avalanche_1999,hinrichsen_non-equilibrium_2000}.
At the same time, some studies of neural activity reported exponents deviating from the  directed percolation universality class~\cite{fontenele_criticality_2019,mariani_disentangling_2022}.
Taken together, network topology should be considered when interpreting and modeling the empirical avalanche statistics.

To investigate the relationship between network topology and critical dynamics, we developed a finite-size branching network model with various connectivity structures, ranging from spatially arranged networks resembling the local connectivity structure of the cortex to random or all-to-all connectivity.
Using this model, we show how the network topology affects the critical branching parameter, avalanche-size distributions, and their critical exponents.
We find that the non-critical avalanche-size distributions can  appear as critical in finite networks, so a quasi-critical (i.e. not truly critical but expressing some amount of scaling) finite network can be confused with a mean-field critical one.

The model consists of binary units positioned on a two-dimensional ($L \times L$) square lattice with periodic boundary conditions. 
The connectivity is defined by two parameters: connectivity radius $\rad$ and the probability of rewiring $\prw$.
First, each unit is connected to all units in its $\rad$-Moore neighborhood ($(2\rad+1)^2-1$ neighbors). Then, with probability $\prw$, each connection can be selected for rewiring and then rewired to a uniformly chosen random location.
To study the impact of different network topologies on the dynamics, we consider a network with $\rad = 1$ and $\prw = 0$ and then systematically increase either the radius (\fig\ref{fig:model_conn}, top) or the rewiring probability (\fig\ref{fig:model_conn}, bottom). 
For $\rad = L/2$, a network with size $N=L^2$ will be all-to-all connected, without any structure.
For $\prw = 1$, we obtain a completely randomly connected network. 
Both limit cases correspond to the usual mean-field configuration.

Each unit $i$ in the network transitions stochastically between an active state $s_i = 1$ and an inactive state $s_i=0$, depending on the connectivity and external input~\footnote{Codes for simulating the network model and replicating the results are available both in Python and C++ at \href{https://github.com/LevinaLab/IBNM}{https://github.com/LevinaLab/IBNM}.}:
\begin{eqnarray}
    \label{equ:trans_prob}
    && p(s_i = 0 \to 1) = 1 - (1-\pext) (1-\pnei)^{\sum_{j\in\Omega_i} s_j}, \\
    && p(s_i = 1 \to 0) = (1-\pext) (1-\pself) (1-\pnei)^{\sum_{j\in\Omega_i} s_j},
\end{eqnarray}
where $\pnei$ is the probability to be excited by the active neighbor, $\pext$ is the probability to receive external input, $\Omega_i$ represents the set of neighbors of the $i$-th unit, and $\pself$ is the probability to maintain the active state. 
This model is inspired by the activity and interactions in the cortex: the units represent cortical columns, and the active and inactive states correspond to transient high and low levels of activity found in primate visual cortex~\cite{engel_selective_2016, vankempen_top-down_2021}. 
The probability $\pself$ accounts for recurrent interactions between neurons within one column and $\pnei$ represents  recurrence between columns (more details in~\cite{zeraati_a_2021,shi_spatial_2022}). For simulations, we assume there is no external input ($\pext=0$) and take $\pself = 0.5$.  

The model is a spatially structured version of a branching network (BN)~\cite{haldeman_critical_2005, kinouchi_optimal_2006}. 
The BN with random connectivity is completely described by local branching parameter $\bp = \pself + |\Omega_i| \pnei$, summing all the outgoing connection probabilities of one node. 
On average, the number of active units $A_t$ at time $t$ when there was a single active unit at the previous time step is $\EX{A_{t} | A_{t-1} = 1} = \bp$. By taking $\pnei = (\bp- \pself) / \langle |\Omega_i|\rangle$ and taking $\pself = 0.5$ we re-parameterize the model in terms of $\bp$. 

We first analyze the location of the critical transition depending on network topology. 
The local branching parameter $\bp$ is used as the control parameter. 
The exact location of the critical transition can be found by several methods (\suppsec\ref{app:critical}). 
Here, we look for the critical branching parameter $\bp_c$ that maximizes the variance of the activity $\chi(\rho(\bp))$.
In the mean-field system, the absorbing-active transition happens at $\bp_c = \bp_{MF}=1$ \cite{henkel_non-equilibrium_2008a}.
However, we find that the location of the critical transition depends on the network topology, in agreement with the theory of critical phenomena~\cite{munoz_avalanche_1999, binney_the_2001} (\fig\ref{fig:phase_diag}). 
For the structured connectivity ($\rad = 1$), the phase transition happens at a larger critical branching parameter ($\bp_c = 1.109$). 
As we move towards the mean-field connectivity (either all-to-all, or random), the critical branching parameter gradually converges to $\bp_{MF} = 1$ (\fig\ref{fig:phase_diag}). 
In the structured networks, we refer to the dynamics at the mean-field branching parameter ($\bp_{MF}=1$) as quasi-critical. We will demonstrate in the following that the quasi-critical dynamics presents the classical mean-field scaling with an exponent close to $1.5$ but only for the limited range of the event sizes. Thus, $\bp_{MF}=1$ is not an actual critical point in structured networks.

The location of $\bp_c$ in structured networks is model-dependent and not a universal feature. For example, in the two-dimensional contact process (CP) the critical point is located at $\bp_\textrm{CP} \approx 1.6$ \cite{henkel_non-equilibrium_2008a}. 
In contrast, in our structured model ($\rad = 1$) criticality appears around $\bp_c \approx 1.1$ (see Appendix \ref{app:mf-avalanches} for a mapping between our model and a continuous time CP model).

 \begin{figure}[!t]
     \centering
     \includegraphics[trim=0 0 0 0, clip, width=\columnwidth]{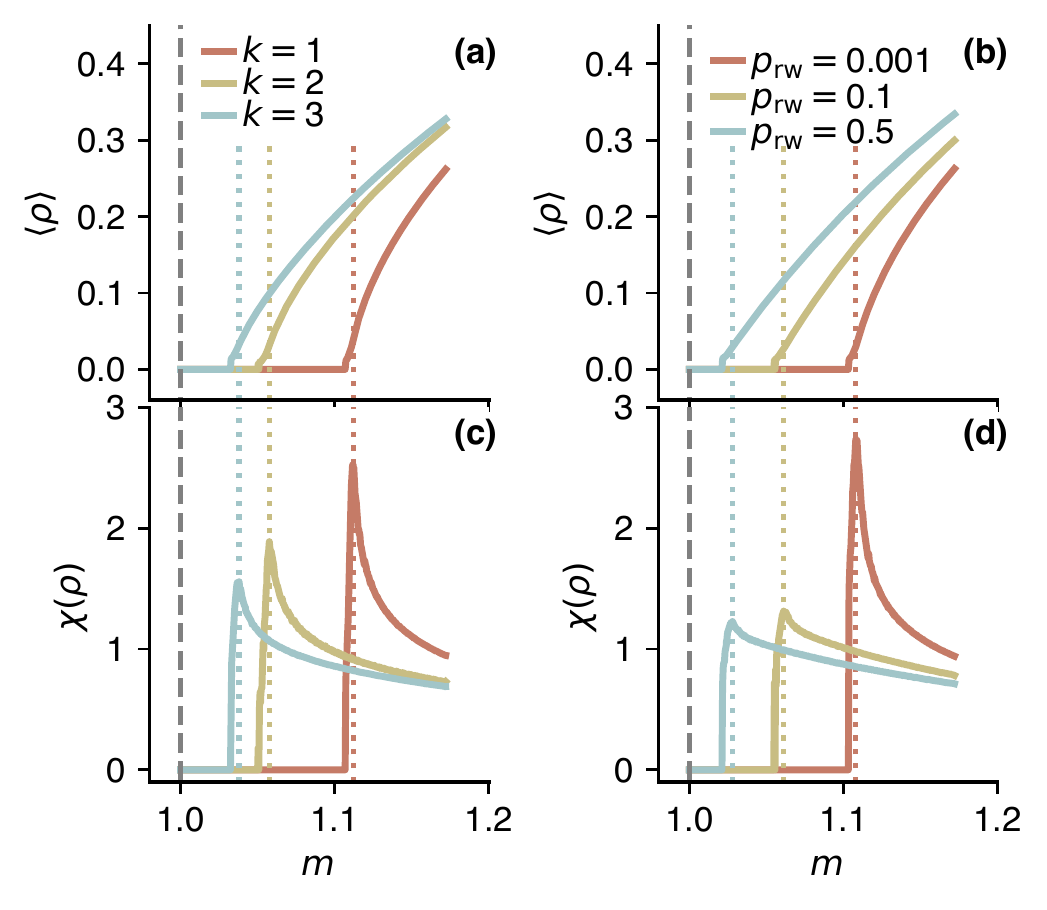}
     \caption{ Location of the critical transition depends on the network topology.
     Criticality occurs at the transition to the non-zero mean activity $\langle \rho \rangle$ (a, b) and maximal variance $\chi(\rho)$ (c, d), represented by vertical dotted lines. With increasing connectivity radius $\rad$ (a, c) or rewiring probability $\prw$ (b, d), the critical branching parameter moves to the mean-field value (vertical gray dashed line). For simulations $L=128$.} 
     \label{fig:phase_diag}
 \end{figure}

Next, we compare the avalanche-size distributions---often seen as the primary indicator for critical behavior in neural data---between the quasi-critical (with $\bp_{MF}$) and critical (with $\bp_{c}$) networks for various network structures and sizes.
An avalanche is a cascade of activity propagation in the network. It starts with an external input activating a single neuron in a quiescent network and ends when the activity dies out. %between quiescent states.
At criticality, avalanche sizes and durations follow a power-law distribution with an exponential cutoff whose location scales with the network size.

Quasi-critical networks exhibit apparent scale-free avalanche-size distributions with the expected MF-DP power-law exponent ($\tau \approx 1.5$, \fig\ref{fig:avs-k}, top). 
In particular, small quasi-critical networks with a finite interaction radius  (e.g., $k=3$ and $L = 8, 32$) can seemingly display finite-size scaling. 
The apparent scaling is more visible in networks with larger connectivity radius $k$ (\suppfig \ref{fig:avalk5}). 
However, for sufficiently large system sizes a characteristic scale becomes evident (cutoff stays independent of the system size when $L>L_\mathrm{scale}$).
The characteristic scale in quasi-critical networks becomes more apparent when compared to the avalanche-size distributions of critical networks (with branching parameter found in \fig\ref{fig:phase_diag}) with the same size and topology (\fig\ref{fig:avs-k}, bottom). 
In critical networks, power laws extend up to much larger sizes. For nearest-neighbors connectivity the exponent shifts from the mean-field values of $\tau\approx1.5$ towards $\tau\approx1.27$ as expected for the two-dimensional directed-percolation (2D-DP) universality class ~\cite{munoz_avalanche_1999} (they approach $\tau=1.27$ as $N\to\infty$, \suppfig\ref{fig:scaling-large}, \suppsec\ref{app:dimension}).
Changing the network topology towards random or all-to-all connectivity brings the critical point closer to $\bp_{MF}=1$ (\fig\ref{fig:phase_diag}). Hence, with larger $k$, the characteristic scale of avalanche sizes in quasi-critical networks increases and the exponents shifts towards $1.5$ (\fig\ref{fig:avs-k}, \suppfig\ref{supfig:coal_diffNet}).

\begin{figure}[t]
     \centering
      \includegraphics[trim=0 0 0 0, clip, width = 1\linewidth]{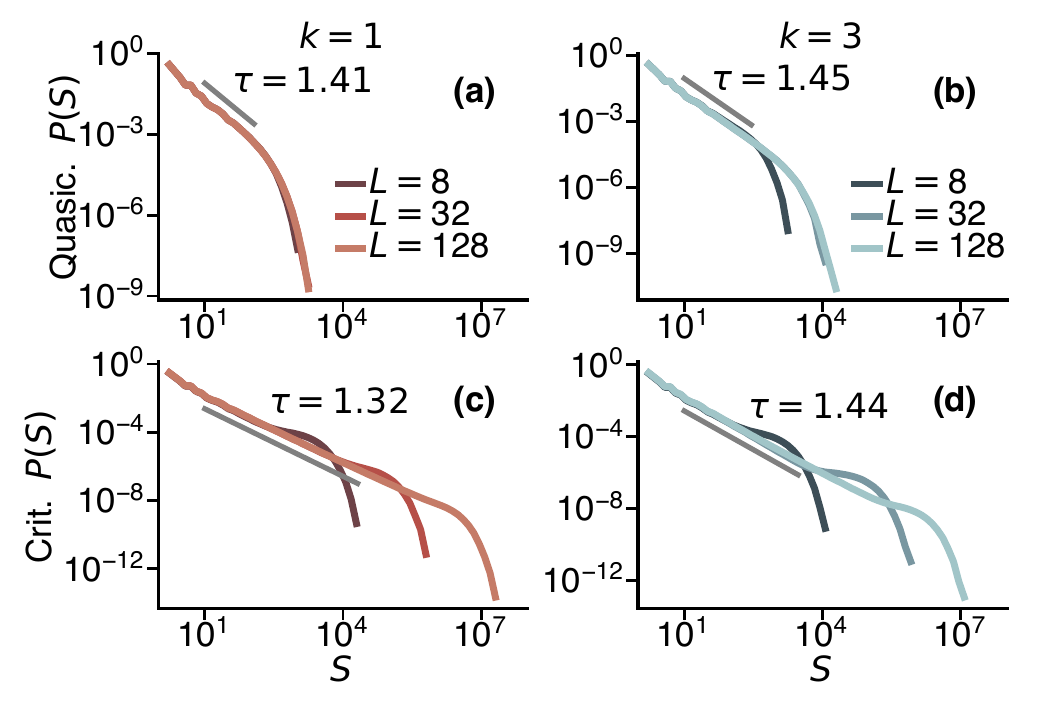}
     \caption{Avalanche-size distributions differ in quasi-critical and critical networks.
     System-size dependence at quasi-criticality (a, b) and criticality (c, d), for $\rad =1$ (a, c) and $\rad =3$ (b, d).
     For critical systems, the cutoff location of the avalanche-size distribution shifts with system size ($L^2$) (c,d). Quasi-critical avalanches follow a power law up to a cutoff that is scaling with the system size for small systems (see also \suppfig\ref{fig:avalk5}) but exhibits a characteristic scale for large systems (a,b). Gray lines indicate the fitted power-law distribution with the exponent $\tau$ for $L = 128$ (see \suppsec\ref{app:av-fit} for fitting details). Critical branching parameters are defined by maximum variance point for each system size, as in \fig\ref{fig:phase_diag}.}
     \label{fig:avs-k}
 \end{figure}

Increasing the connectivity radius or rewiring probability in critical networks changes the critical exponents continuously from $\tau\approx1.27$ (the 2D-DP universality class) to $\tau \approx 1.5$ (the MF-DP universality class).
However, these two mechanisms affect the critical exponents in different ways.
The major difference stems from their behavior in the thermodynamic limit.
For any finite connectivity radius $\rad$ and no rewiring, in the limit of large network sizes, connections are short-ranged and the dynamics belong to the 2D-DP universality class with the critical exponent of $\tau = 1.27$.
At the same time, finite networks with large enough $\rad$ are almost fully connected, showing exponents similar to MF-DP. 
Thus, for fixed $\rad$, the network size affects how close the system is to a fully connected system.
The combination of these two factors leads to the true scaling exponent (known for the 2D-DP and MF-DP~\cite{munoz_avalanche_1999}) being visible only for very large avalanches ($S\gg 1$), which require very large system sizes ($L\to\infty$). In large networks, the power-law exponent will change slowly and continuously from MF-DP for relatively small avalanches to the 2D-DP exponent for large events (\suppfig\ref{fig:scaling-large}).
The rewiring, on the other hand, conserves the network topology independently of the system size. 
Therefore, the critical exponents increase with increasing rewiring probability, approaching the mean-field limit.
For a fixed $\prw$, avalanche statistics follows a well-defined power-law distribution with a single exponent, which displays the true scaling also for small event sizes and durations.
The difference between both cases can be clearly understood in how the effective dimension of the network changes with increasing $k$ or $\prw$ (\suppsec\ref{app:dimension}).

Our results suggest that observing the power-law-like behavior in avalanche-size distributions from small networks cannot be a reliable signature of criticality even when the distributions scale with the system size.
Close to criticality, subcritical systems can exhibit apparent scale-free behavior, where the characteristic scale is only uncovered in the limit of large systems.

We next investigate the mechanisms underlying the differences between the avalanche dynamics in mean-field and structured networks.
Due to locally structured connectivity in our model, each unit can be activated by multiple sources at the same time, generating \textit{coalescence} (\fig\ref{fig:coal}a). 
We measure the network coalescence from the ongoing activity in the timescale-separated regime ($\pext = 0$), where each new avalanche is initiated after the previous one is over.
Each unit $i$ in the network can be simultaneously activated by multiple neighbors, or by the self-excitation (\fig\ref{fig:coal}a). We define the local coalescence as the number of sources that activated unit $i$ minus one. 
Let $n_{i,t}$ be a number of active neighbors of unit $i$ at time $t$. The local coalescence of unit $i$ at time $t$ is a random variable 
\begin{equation}   
    \label{equ:coal_local}
\coal_{i,t} = \max \left(0, \sum_{k=1}^{n_{i,t}} b_k + s_i \cdot b_0 - 1 \right ), 
\end{equation}
where $b_k \sim \mathrm{Bernoulli} (\pnei) $ and  $b_0 \sim \mathrm{Bernoulli} (\pself)$. 
Let $A_t$ be the number of active units in the network at time $t$. 
Then, the average normalized network coalescence $\coal(A)$ is given by 
\begin{equation}   
   \label{equ:coal}
    \coal(A) = \frac{1}{A} \left\langle \left. \sum_{i=1}^{L\times L}\coal_{i,t} \right \vert A_t =A \right\rangle,
\end{equation}
where average is taken over all the times with $A_t =A$.

Due to coalescence, the effective branching parameter is smaller than the local branching parameter and depends on the network topology and the number of active units.
We can estimate the effective branching parameter for the given number of active units $A$ from the simulated activity as:
\begin{equation}   
    \label{equ:bp_effective}
    \bp_\textrm{eff} (A)  = \bp - \coal(A).
\end{equation}
The estimated effective branching parameter satisfies $\bp_\textrm{eff}(A) < \bp$ but varies depending on the number of active units $A$ (\fig\ref{fig:coal}b,c, left). 
For small $A$, $\bp_\textrm{eff}(A)$ values are closer to $\bp$, but with increasing number of active units, the coalescence also increases leading to larger deviations of $\bp_\textrm{eff}(A)$ from $\bp$. 
Therefore, a larger local branching parameter ($\bp = 1.109$) is required to have an effective branching parameter close to $1$ that creates critical dynamics. 
When increasing the connectivity radius or the rewiring probability, the activity can spread to a broader range of units. In addition, increasing the connectivity radius while keeping the branching parameter constant reduces the strength of individual connections, and coalescence for every unit scales as a power of the connection strength. Thus, by increasing the connectivity radius or the rewiring probability, the coalescence decreases, and the $\bp_\textrm{eff}(A)$ becomes closer to $\bp$ (\suppfig\ref{supfig:coal_diffNet}, consistent with \fig\ref{fig:phase_diag}).

To capture the impact of coalescence on avalanche-size distributions, we simulated an equivalent adaptive branching process for each network model.
To generate this process, we first find the $ \bp_\textrm{eff} (A)$ from a long simulation of the network model using Eq.~(\ref{equ:bp_effective}). 
Then we define the adaptive branching process, as a Markov process $\tilde{A}(t)$, where each of the ancestors can generate a binomially distributed number of offsprings $z_i$,   
\begin{equation}   
    \label{equ:bprocess}
   \tilde{A}_{t+1} = \sum_{i=1}^{\tilde{A}_t} z_i(\tilde{A}_t), \quad z_i \sim \mathrm{Bin} \left(n, \frac{\bp_\textrm{eff} (\tilde{A}_t)}{n} \right), 
\end{equation}
$n$ is a maximal number of offsprings for one ancestor that is equal to the number of neighbors $(2k+1)^2-1$ in the original network. For the simulations and figures in the paper we took $n=8$.

\begin{figure}[t]
    \centering
     \includegraphics[trim=0 0 0 0, clip, width = 1.\linewidth]{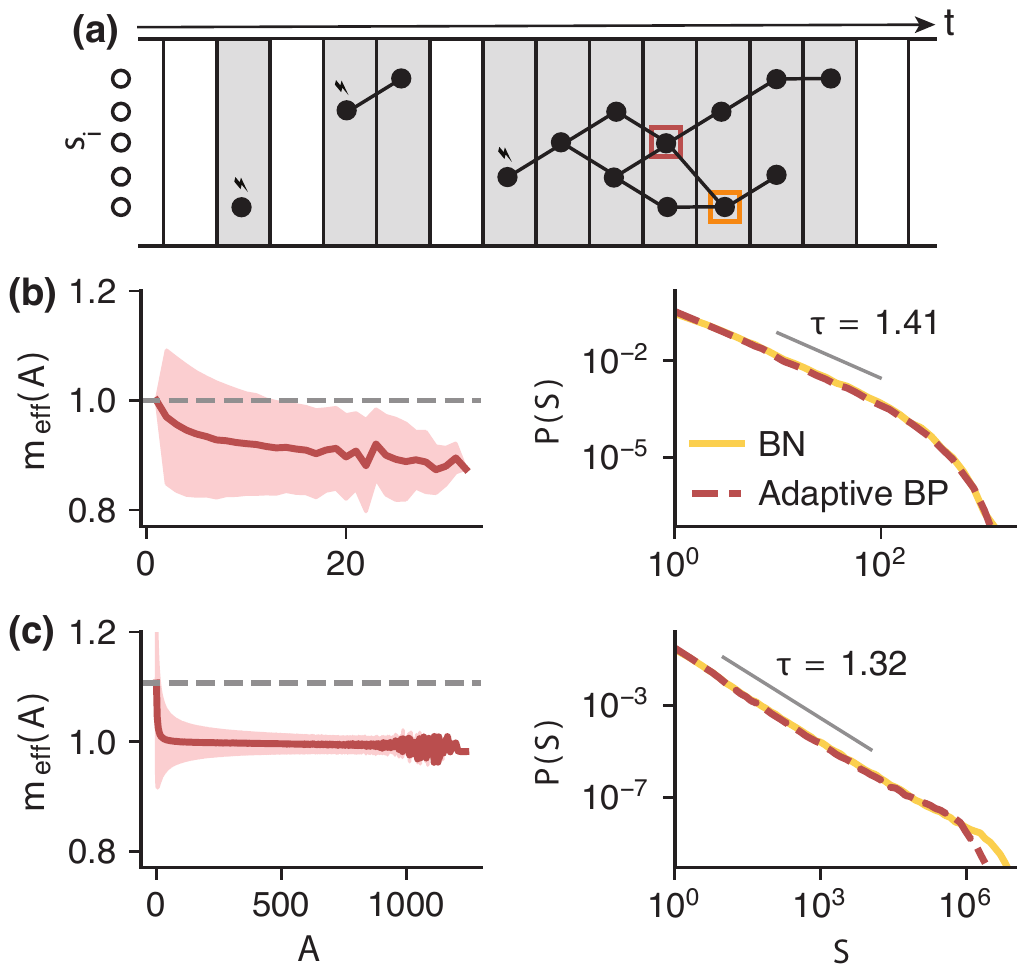}
    \caption{{The adaptive branching process 
    captures the shape of the avalanche-size distributions in structured networks ($\rad = 1$, $L = 128$).} 
    (a) Avalanches (gray frames, each row marked with an empty circle represents one network unit, filled circles are active units) are separated by quiescent moments (white frames). Two types of coalescence: simultaneous activation by multiple active neighbors (brown square), or by the self-excitation and an active neighbor (orange square). Each avalanche starts with the external input (black lightning bolts). The arrow indicates time.
    (b, left) The effective branching parameter ($\bp_\textrm{eff}(A)$, brown line) in the quasi-critical ($m = 1$) network as a function of the number of active units ($A$, Eq.~(\ref{equ:bp_effective})) deviates from the local branching parameter (dashed line). The shading indicates $\pm 1$ s.d. of $C_{i,t}|A$.
    (b, right) The adaptive branching process (dashed brown line) has the same avalanche-size distribution ($P(S)$) as the structured branching network (BN, yellow line). The gray line shows the power-law fit with the exponent $\tau$.
    (c) Same as (b) for the critical network ($m = 1.109$).}
    \label{fig:coal}
\end{figure}

Avalanche-size distributions of the adaptive branching processes well approximate the shape of corresponding distributions from the network dynamics (\fig\ref{fig:coal}b,c, right). 
In the quasi-critical network, distributions for the branching process and the network are completely overlapping. 
The adaptive branching process for the critical network captures most of the distribution, except for some deviations in the tail, and it has a similar power-law exponent.
The mismatch in the tail can be generated by a too scarce sampling of large avalanches to estimate the correct effective branching parameter.
Overall, despite the large variability in the network coalescence, the average value of effective branching parameter is sufficient to predict the shape and exponent of the avalanche-size distribution in the structured networks.

The network coalescence can be estimated analytically for a branching network with random connectivity~\cite{zierenberg_description_2020}. 
However, analytical determination of coalescence for finite-dimensional topologies (e.g., structured networks) requires renormalization group approaches relying on the precise knowledge of the system dimension \cite{hinrichsen_non-equilibrium_2000,henkel_non-equilibrium_2008a}. This approach becomes especially difficult when dealing with finite system sizes.
Hence, we used a simulation based approach to estimate the coalescence from the network activity.

Our results indicate that the differences between the dynamics in structured and mean-field networks arise from the coalescence created by the local network interactions. 
Increasing the connectivity radius or rewiring probability reduces the coalescence and the network dynamics becomes more similar to the conventional branching process.
Dependence of coalescence on network topology is in line with the previous observation that the effective branching parameter reduces with increasing degree of network clustering~\cite{keeling_implications_2005}. 
Although there have been many studies addressing the effect of long-range connections in the brain \cite{valverde_structural_2015}, and critical-like avalanches have been reported in small-world \cite{ferreira_critical_2013} and scale-free \cite{dearcangelis_self-organized_2002} networks, there was no systematic study of the impact of varying coalescence on avalanche distributions.

We demonstrated how a range of scaling exponents can arise from the network structure and be misinterpreted in finite-size networks.
This is particularly significant for interpreting observations from finite-size neural recordings, where the number of recorded neurons or electrodes are often treated as a proxy for system size, and finite-size scaling is assessed by down-sampling the recorded neurons (electrodes)~\cite{yu2014scale}. The precise number of neurons involved  in the dynamics and their interaction radius is often unknown, and approaches which can recover the dynamical regime from subsampled networks~\cite{levina_subsampling_2017a} so far mainly rely on the down-sampling of available data.

We showed that while increasing the connectivity radius and rewiring affect the dimensionality of the network in different ways, both mechanisms can create exponents in between the 2D-DP universality class and the mean-field values.
At the same time, critical exponents other than the MF-DP have been recently observed in neural activity~\cite{fontenele_criticality_2019} and linked to a phase transition at the onset of collective oscillations in a network model with excitatory and inhibitory neurons~\cite{fontenele_criticality_2019,porta_modeling_2019}.
Moreover, it was shown analytically that the absence of separation of timescales in measuring avalanches could lead to exponents between directed and standard percolation universality classes~\cite{korchinski_criticality_2021}, or in a particular case  to $1.25$~\cite{das2019critical}.
How mixture of different mechanisms affects
observed avalanche statistics should be studied in the future to better understand the underlying mechanisms for variable exponents in neural activity.

This work was supported by a Sofja Kovalevskaja Award
from the Alexander von Humboldt Foundation, endowed by the Federal Ministry of Education and Research (RZ, VB, AL), SMARTSTART2 program provided by Bernstein Center for Computational Neuroscience and Volkswagen Foundation (RZ), NIH grants RF1DA055666 (TAE),  the Sloan Research Fellowship from the Alfred Sloan Foundation (TAE). Simulations were performed with assistance from the NIH Grant S10OD028632-01. We acknowledge the  support from the BMBF through the T\"ubingen AI Center (FKZ: 01IS18039B) and International Max Planck Research School for the Mechanisms of Mental Function and Dysfunction (IMPRS-MMFD). We also thank Miguel A. Mu\~noz and Giorgio Nicoletti for valuable discussions.


%%%%%%%%%%%%%%%%%%%%%%%%%%%%%%%%%%%%%%%%%%%%%%%%%%%%%%%%%%%%%
%%%%%%%%%     Supplementary materials     %%%%%%%%%%%%%%%%%%%%%
%%%%%%%%%%%%%%%%%%%%%%%%%%%%%%%%%%%%%%%%%%%%%%%%%%%%%%%%%%%%%
\clearpage
\renewcommand\thefigure{\arabic{figure}} 
\renewcommand{\figurename}{Supplementary Fig.}
\renewcommand{\tablename}{Supplementary Table}
\setcounter{figure}{0}  
\setcounter{table}{0}

\title{Supplementary information for: Topology-dependent coalescence controls power-law exponents in finite networks}
\maketitle

%______________________________
\appendix
\section{Estimating the critical branching parameter using phase diagrams}
\label{app:critical}
%______________________________

The critical branching parameter has been estimated by computing the location of the maximum activity susceptibility. 
Estimation of the critical transition's location for absorbing-active phase transitions is complicated since at criticality the stationary state is the absorbing one. 
Hence, one is forced to make statistics over pseudo-stationary states \cite{hinrichsen_non-equilibrium_2000, henkel_non-equilibrium_2008a}. 
Phase diagrams are computed by letting the simulations run for a fixed time $t_{sim}=10^5$, which is set to be as large as possible. 
The first two moments of the total particle density are measured during this time, letting enough time separation between measurements to avoid any correlation bias. 
These allow us to obtain the average density $\langle \rho \rangle$ and the susceptibility $\chi (\rho) = \sqrt{N} \left( \langle \rho^2 \rangle - \langle\rho\rangle^2 \right)$.
If the simulation falls into the absorbing state before reaching $t_{sim}$, measurements are discarded (setting mean and variance equal to zero) and the procedure starts again, so the results are averaged only over runs that survived.
Near criticality, even when activity eventually falls to zero, a small density is still able to produce avalanches at any time, so there is always a non-vanishing probability of observing any amount of activity at $t_{sim}$, which grows with system size.
This is different from the active phase, where activity can be arbitrarily small, but fluctuates around its mean value.  

Then, one computes the average density and its susceptibility, looking for the largest susceptibility to have an estimation for the location of the critical transition.
To get a better estimation for the critical control parameter, the usual technique is finite-size scaling, computing the average density for survived runs for long simulation durations and increasing sizes. 
Criticality fulfills power-law decay of the density $\rho^{-N}$, while the active phase saturates and the subcritical decays exponentially. 
However, this method requires very long simulation times, large system sizes, and many runs -in order to have good statistics over the survival ones-, which in our model lead to very long computation times~\footnote{Other models such as the contact process can be simulated more efficiently due to their asynchronous nature, and the fact that active individuals do not interact anymore with other active individuals.}. 

Finally, in contrast with continuous-time models, here we use a non-linear probabilistic model simulated in discrete times, leading to a more complex behavior for the transition rates than in the classical contact process. Hence, the exact location of the critical point needs a very accurate determination of the recurrent probability $p_r$, as demonstrated in \suppsec\ref{app:mf-avalanches}.

%______________________________
\section{Estimating the power-law exponent from the avalanche-size distributions}
\label{app:av-fit}
%______________________________
 We compute the avalanche-size distributions in the separated timescale regime ($\pext = 0$). Each avalanche starts with a single active unit and ends when the whole network activity dies out. We define the size of an avalanche as the total number of units activated during the avalanche. To obtain the avalanche-size distributions we simulated $10^7$ avalanches for each network. For coalescence analysis and the adaptive branching process we simulated $10^5$ avalanches.

At the critical point, the size of avalanches $\av$ follows a power-law distribution. 
We estimate the power-law exponent $\expo$ by fitting the avalanche-size distribution with a discrete and truncated power-law distribution as~\cite{clauset_power-law_2009, klaus_statistical_2011}
\begin{equation}  
    \label{equ:power_law}
    P(\av) = \frac{\av^{-\expo}}{\zeta (\expo , \av_{min})- \zeta (\expo , \av_{max})}.
\end{equation}
Here, $\av_{min}$ and $\av_{max}$ are, respectively, the minimum and maximum avalanche size considered for fitting and  $\zeta (\expo , \av)$ is the Hurwitz zeta function defined as,
\begin{equation}   
    \label{equ:zeta_function}
    \zeta (\expo , \av) = \sum_{n=0}^{\infty}(n+\av)^{-\expo}.
\end{equation}
We find the optimal value of $\expo$ using the Maximum likelihood estimation (MLE) with a grid search. 
For the fits, we set $\av_{min} = 10$ and $\av_{max}$ to $96$ percentile of the distribution. 

\begin{figure}[!t]
     \centering
  \includegraphics[trim=0 0 0 0, clip, width=\columnwidth]{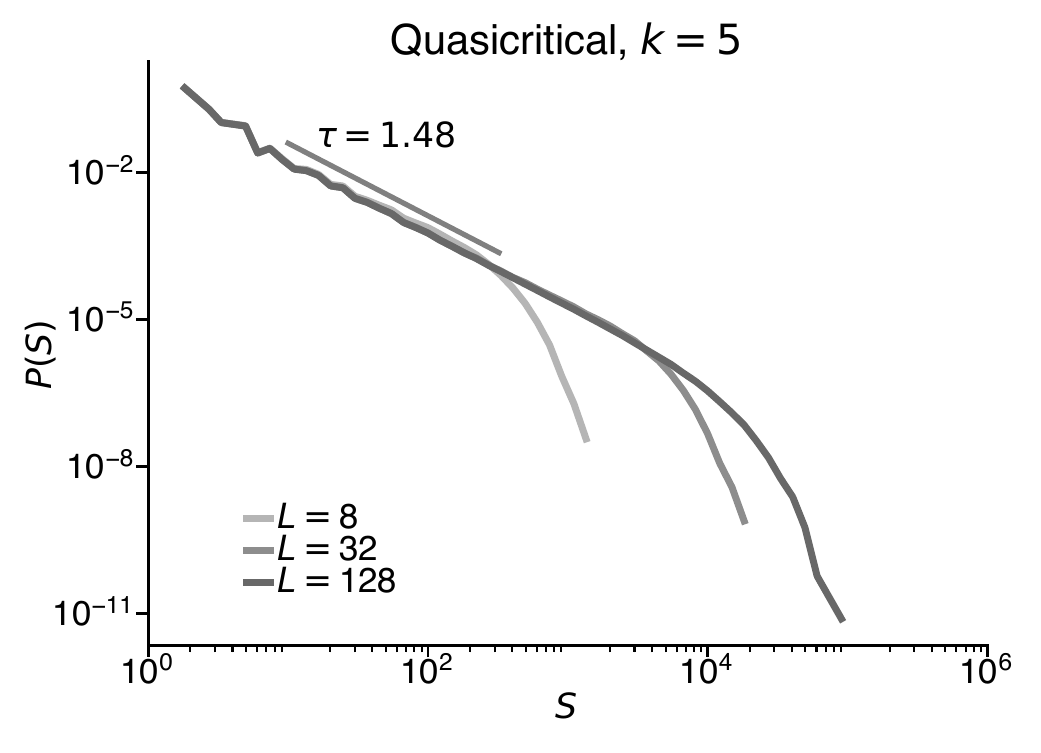}
     \caption{ Quasi-critical ($\bp=1$) avalanche size distribution for $k=5$. At finite sizes, avalanches for larger $k$ present seemingly critical distributions, with a cutoff shifting with the system size and the mean-field exponent close to $1.5$.} 
     \label{fig:avalk5}
 \end{figure}

We found that the estimated exponents depend on the network topology (\fig\ref{fig:avs-k}). In particular,  \suppfig\ref{fig:avalk5} shows the avalanche-size distribution for $k=5$ quasicritical network, which has an exponent closer to $1.5$ than the low-k cases shown in the main text (\fig\ref{fig:avs-k}b,c).  
This is because the critical local branching parameter $m_c$ approaches $1$ as $k$ increases for fixed $N$, meaning that the quasicritical network becomes closer to actual criticality with a larger $k$. We also observe an apparent shift of the cutoff with the system size (that will disappear for even larger networks).
Therefore, finite structured quasi-critical networks can appear critical when the connectivity radius is large enough \suppfig\ref{fig:avalk5} or the network size is small (\fig\ref{fig:avs-k}b,c).
 
%______________________________
\section{Growing radius vs rewiring: the effect of dimensionality}
\label{app:dimension}
%______________________________
We argued in the main text that increasing both the connectivity radius and rewiring probability brings the critical exponents closer to the mean-field ones, while exhibiting different behaviors in the thermodynamic limit behavior. 
In this Appendix, we show how the topology modulates the dimension of the system. Dimensionality plays an essential role in the theory of critical phenomena \cite{hinrichsen_non-equilibrium_2000, henkel_non-equilibrium_2008a, binney_the_2001}.
One can see that in the increasing radius case the critical exponents must be continuously varying based on the following argument: for any arbitrarily large size $N$, one can always take a large enough $k$ to make the network almost fully-connected, displaying avalanches with near-mean-field scaling, up to size $S_\textrm{max}(N)$ which depends on $N$. For $S > S_\textrm{max}$, $P(S)$ is no longer described by a power-law due to finite-size effects. Now, if $N$ is further increased to $N'\geq N$, the distribution up to $S_\textrm{max}(N)$ must be exactly the same as before, since by hypothesis we assumed all boundary effects are taking place for larger sizes. At the same time, we know that for a critical system we have finite-size scaling, so the scale-free distribution now must hold for $S_\textrm{max}(N) < S < S_\textrm{max}(N')$. But if $N'\to\infty$, then $k\ll N'$ again, making the network structured so the exponent of this new scaling should be different to the one presented until $S_\textrm{max}$.

 \begin{figure}[!t]
     \centering
  \includegraphics[trim=0 0 0 0, clip, width=\columnwidth]{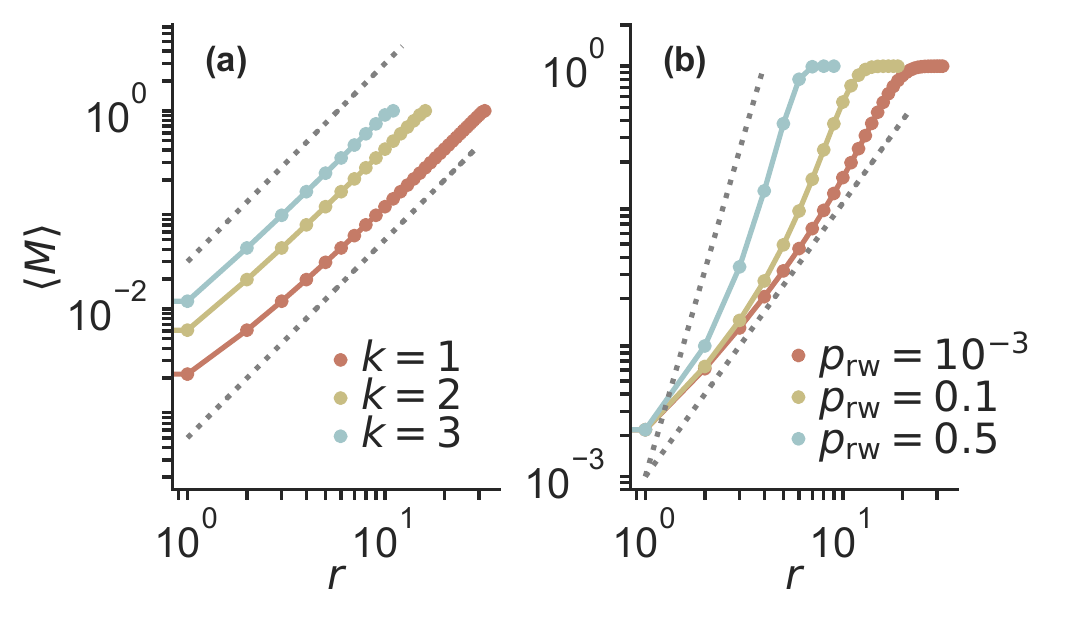}
     \caption{ Dimensionality estimation based on topology.
     (a) Scaling of $\langle M \rangle$ --the average total number of nodes at distance less than or equal to $r$-- as $r$ increases for different connectivity radii $k$. Notice that the slope is always the same, $d=2$, marked with dashed black lines, but as $k$ grows, the saturation appears in smaller $\rad$. (b) Same as a for increasing rewiring probability $\prw$. The saturation point changes due to different slopes, increasing with $\prw$. Dimensions $d=2$ and $d=5$ are indicated for reference (dashed back lines).} 
     \label{fig:dimensionality}
 \end{figure}

\begin{figure}[!t]
     \centering
  \includegraphics[trim=0 0 0 0, clip, width=\columnwidth]{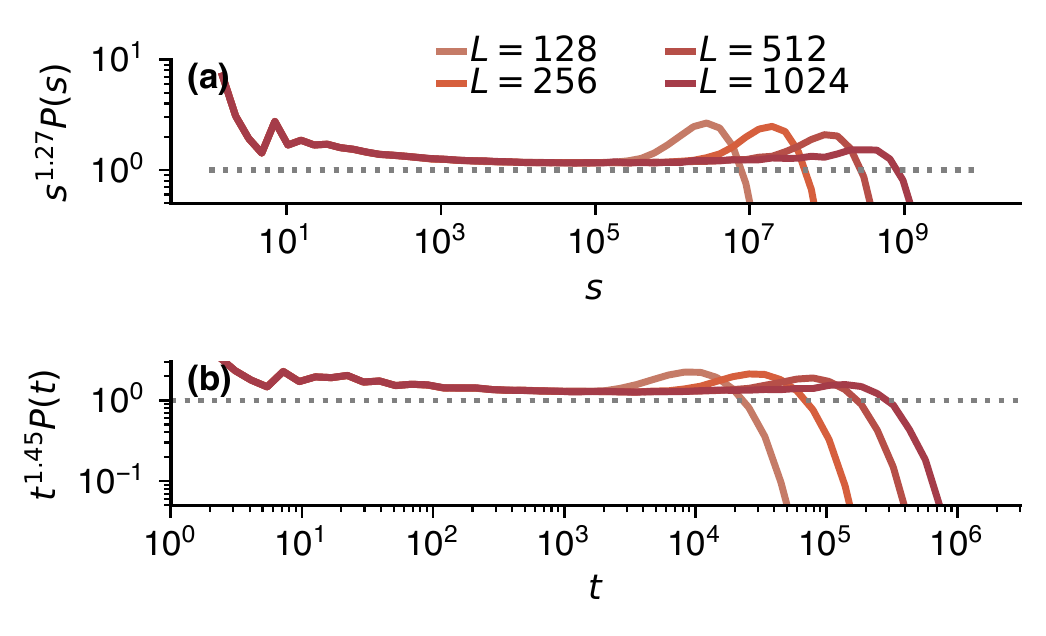}
     \caption{True scaling of the system in the thermodynamic limit at criticality for $k=1$. Avalanches of very large system sizes are plotted to show that as $N\to+\infty$ the expected 2D directed percolation exponents ($\tau=1.27$ for sizes and $\alpha=1.45$ for duration \cite{munoz_avalanche_1999}) are recovered. Avalanche-size (a) and duration distributions (b), scaled to render distributions horizontally based on the 2D directed percolation exponents (dashed lines) to appreciate the possible deviations from the theoretical scaling in detail. Parameters, $p_s=0$, $p_r=1.08975$, and $10^7$ avalanches.} 
     \label{fig:scaling-large}
 \end{figure}

One could argue that by increasing $N'$ the network effective dimensionality is reduced to $d=2$. In fact, the classical definition of dimension is that the mass encompassed in the ball of radius $r$ scales as $M\sim r^d$. When this scaling relationship is not present, it is assumed that dimension is not well defined (i.e., $d \to \infty$). One can naively generalize the dimension definition to networks, by letting $r$ to be the distance between nodes, and $\langle M \rangle$ the average total number of nodes at distance less or equal than $r$. For the structured lattice, it is clear that $M\sim r^2$, as long as the network is infinite. If the network is finite, then there is a distance $r^*(k,N)$ such that $\langle M(r\geq r^*) \rangle = N$. Hence, the network appears to be 2D for $r\leq r^*$. In practice, even for low values of $k$ one needs huge sizes to see the 2D network scaling. 
\suppfig\ref{fig:scaling-large} shows that indeed the structured system with $k=1$ relaxes to the critical exponent of the directed percolation universality class, $\tau \approx 1.27$.
However, notice that the scaling is not clear until reaching very large system sizes, due to the use of the Moore neighborhood instead of the classical 4-neighbor Manhattan lattice.

\begin{figure}[t]
    \centering
     \includegraphics[trim=0 0 0 0, clip, width = 1.\linewidth]{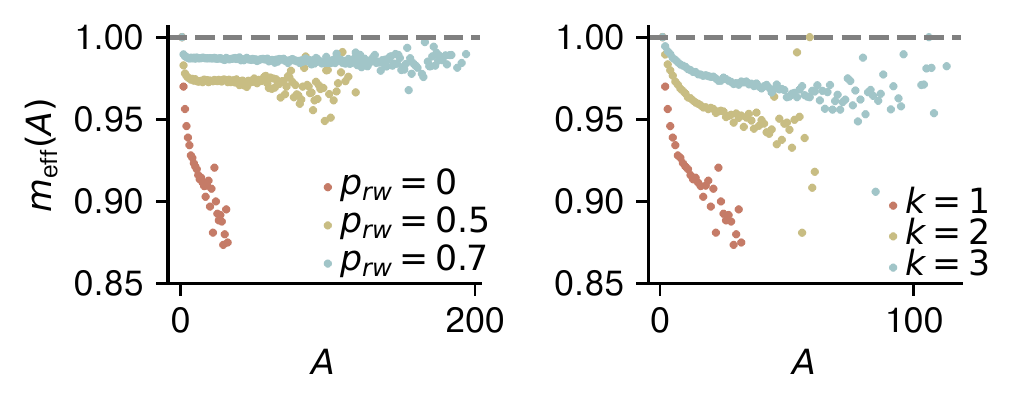}
    \caption{{Dependence of coalescence and effective branching parameter on network topology.} 
    By increasing rewiring probability $\prw$ (left) or connectivity radius $\rad$ (right), the coalescence decreases and effective branching parameter $\bp_\textrm{eff}$ becomes closer to the local branching parameter $\bp$ (dashed line). However, changes in the effective branching parameter differs between the two mechanisms. }
    \label{supfig:coal_diffNet}
\end{figure}

For the case of rewiring, long-ranged connections allow connecting any arbitrary pair of nodes in a small number of steps, making again $\langle M\rangle=N$ for small distances. However, in this case, as rewiring probability $\prw$ is increased, so does the slope of $\langle M(r) \rangle$. In this case, the scaling relation is only lost if the network is completely random, since this is the only case in which any two nodes could be possibly connected at a finite distance. Any small amount of structure will make certain nodes infinitely separated, allowing to fulfill the scaling for $\langle M \rangle$. \suppfig\ref{fig:dimensionality} illustrates the differences in the function $\langle M \rangle$ between the increasing radius and the rewiring cases.
These difference can also be observed when measuring the effective branching parameter from the dynamics of these two network types (\suppfig\ref{supfig:coal_diffNet}).

%______________________________
\section{Continuous mean-field approach}
\label{app:mf-avalanches}
%______________________________ 

It is possible to demonstrate that our discrete model has a second order phase transition in mean-field, and to obtain an exact relationship between the probabilities $p_s$, $p_r$ and the branching ratio $m$. In order to do so, we proceed in the following way: first, the discrete probabilities are written as a continuous Markov process, from which it is possible to derive a Master equation to apply our formalism; second, one performs a Kramers-Moyal expansion of the master equation, from which it is possible to identify a Langevin dynamics for the density of active particles; finally, this equation is expanded near the absorbing state and mapped to the "normal form" of the contact process.

First, under the mean-field approach, the transition probability of a single node becoming active is given by 

\begin{equation}
    p\left(0\to1\right)=1-\left(1-p_{r}\right)^{x}\equiv F\left(x\right),
\end{equation}

where $x=A/N$ is the particle density of the system, which is an intensive variable, well defined in the thermodynamic limit. Thus, the transition probability matrix between states is given by 

\begin{equation}
    \hat{P}\left(\Delta t\right)=\left(\begin{array}{cc}
1-F\left(x\right) & F\left(x\right)\\
\left(1-p_{s}\right)\left(1-F\left(x\right)\right) & p_{s}+\left(1-p_{s}\right)F\left(x\right)
\end{array}\right).
\end{equation}

In a continuous time model described by the Markov transition matrix $\hat Q$, the probability that a transition took place during the timestep $\Delta t$ is given by $\hat P = \exp\left(\hat Q \Delta t \right)$. Equating both allows us to find the Markov transition rates~\cite{shi_spatial_2022},

\begin{align}
    \omega\left(0\to1\right)=&-\frac{F\left(x\right)}{\Delta t\left[1-p_{s}\left(1-F\left(x\right)\right)\right]}\log\left[\left(1-p_{s}\right)F\left(x\right)\right], \\
    \omega\left(1\to0\right)=&-\frac{\left[1-F\left(x\right)\right]\left(1-p_{s}\right)}{\Delta t\left[1-p_{s}\left(1-F\left(x\right)\right)\right]}\log\left[\left(1-p_{s}\right)F\left(x\right)\right].
\end{align}

In mean-field, since all the particles are identical, the probability of increasing the activity by one particle is given by the probability of picking an empty site and performing a transition up. Conversely, the probability of decreasing activity is given by the probability of picking an active particle and transitioning down, i.e.

\begin{align}
    \Omega\left(x\to x+\Delta x\right)=&\left(1-x\right)\omega\left(0\to1\right), \\
    \Omega\left(x\to x-\Delta x\right)=&x\omega\left(1\to0\right).
\end{align}

An interesting theoretical note is to realize that the non-linear activation rate will yield arbitrary powers of the density $x$ when is Taylor expanded. This can be interpreted as having $n$-body interactions, since in simple models with linear rates transitions involving $n$ bodies have rates proportional to $x^n$ (the contact process, for example, only involves up to quadratic term). This is a direct consequence of coalescence, and in practice it means that in the discrete model a particle in contact with two active neighbors can be activated by either one of those or by the effect of both acting together. So, if rates are Taylor expanded around the absorbing state $x=0$,

\begin{equation}
    \Omega\left(x\to x\pm\Delta x\right)=\sum_{k=1}^{+\infty}\lambda_{k} ^\pm x^{k},
\end{equation}

then the rate at which a particle in contact with active two neighbors activates is given by $2\lambda_1 ^+ + \lambda_2 ^+$. 

Once the global rates have been identified one can write a Master equation and expand it using the Kramers-Moyal approximation. Since this is a standard procedure we will skip the technical details, redirecting the reader instead to classic textbooks on the subject~\cite{gardiner2009}. One then can show that a Langevin equation for the density of active particles is given by 

\begin{align}
    \dot x =&  \Omega\left(x\to x+\Delta x\right) -  \Omega\left(x\to x-\Delta x\right) + \nonumber \\ 
    +&\frac{1}{\sqrt{N}} \sqrt{\Omega\left(x\to x+\Delta x\right) +  \Omega\left(x\to x-\Delta x\right) } \xi(t),
\end{align}

where $\xi(t)$ is a Gaussian, delta-correlated white noise. Finally, the rates are Taylor expanded. Following the Landau-Ginzburg theory of critical phenomena,  the critical properties of the transition are to be controlled by the first term that becomes always negative (hence, controlling saturation) \cite{binney_the_2001}. It is possible to show that it is sufficient to expand the equation up to second order, 

\begin{equation}
    \dot x =  a_1 x - a_2 x^2 + \sigma \sqrt{x} \xi(t),
\end{equation}

where 

\begin{align}
    a_1 =& \frac{1-p_{s}-\log\left(1-p_{r}\right)}{1-p_{s}}\log p_{s}, \\
    a_2 =& -\frac{\log\left(1-p_{r}\right)}{2\left(1-p_{s}\right)^{2}}\times \\
    \times & \left[2\left(1-p_{s}\right)^{2}+\log\left(1-p_{r}\right)\left(2\left(1-p_{s}\right)+\left(1+p_{s}\right)\log p_{s}\right)\right]. \nonumber 
\end{align}

 \begin{figure}[!t]
     \centering
  \includegraphics[trim=0 0 0 0, clip, width=\columnwidth]{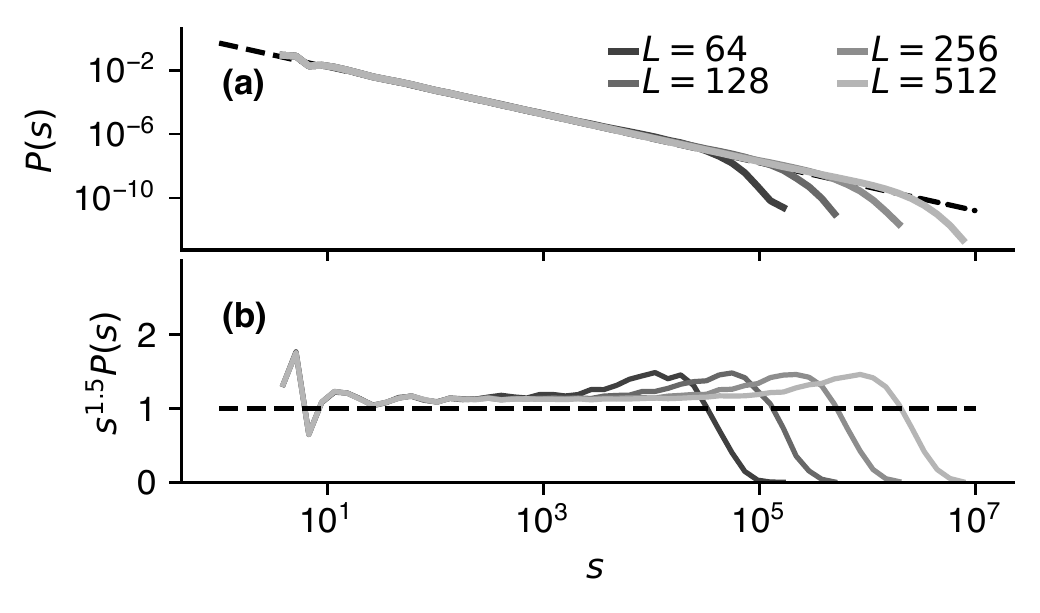}
     \caption{Scaling of mean-field avalanches.
     (a) Finite size scaling of the mean-field avalanche size distribution for different system sizes at criticality, $p_s=0.5$ and $p_r=0.393469$. Theoretical scaling $P(s)\sim s^{-3/2}$ is displayed with a discontinuous line. (b) The same distribution, multiplied by $s^{3/2}$ to display it as a horizontal line, in order to ease the visual inspection of the correct scaling. The distributions were obtained with $10^7$ avalanches.} 
     \label{fig:mfavals}
 \end{figure}

One can demonstrate that $a_2 > 0$ always by direct plotting, or more elegantly, by demonstrating that the function is monotonously increasing and its minimum is positive \footnote{This is done by seeing that $a_2(p_s=0)>0$, $a_2(p_s=1)=0$, and then performing derivatives until it is clear that the function is monotonous}. Critical point happens when the linear term (the "mass") vanishes, which happens at $-\log\left(1-p_{r} \right) = 1-p_{s}^*$. Finally, it is possible to evaluate the branching ratio of the continuous model, knowing that in the contact process we have $a_1 = (1-m)a_2$. The actual relation between the branching ratio and the probabilities is, then,

\begin{widetext}
\begin{equation}
    m = \frac{\log\left(1-p_{r}\right)\left[2\left(1-p_{r}\right)^{2}+\log\left(1-p_{r}\right)\left[2\left(1-p_{s}\right)+\left(1+p_{s}\right)\log p_{s}\right]\right]}{2\left(1-p_{s}\right)\log\left(1-p_{r}\right)\left(1-p_{s}+\log p_{s}\right)+2\left(1-p_{s}\right)^{2}\log p_{s}+\log^{2}\left(1-p_{r}\right)\left[2\left(1-p_{s}\right)+\left(1+p_{s}\right)\log p_{s}\right]},
\end{equation}
\end{widetext}

which reduces to $m=1$ when the critical $p_s ^*$ is set. One then can see that small changes to the probabilities translate into non-linear changes to the branching ratio, which makes the model extremely sensitive to the choice of probability when trying to locate the critical point, and makes it difficult to find a clean scaling for the avalanches for the structured networks. 
In the mean-field case, however, the critical point can be found exactly, and avalanches with the expected exponents are found, as shown in \suppfig\ref{fig:mfavals}.

\end{document}